 \documentclass[12pt]{article} 
\usepackage{graphicx}
\usepackage{comment}
\usepackage{setspace}
\usepackage{amsmath}
\usepackage{amssymb}
\usepackage[textwidth=16cm,textheight=22cm]{geometry}
\usepackage{color}
\usepackage{indentfirst}
\usepackage{xspace}
\usepackage{hyperref}
\usepackage{verbatim}
\usepackage{epstopdf}

\usepackage{lineno}

\def\jour#1#2#3#4{{#1} {\bf#2}, #4 (19#3)} 
\def\jou2#1#2#3#4{{#1} {\bf#2}, #4 (20#3)} 
\def\Natr{{Nature }}
\def\PRL{{Phys. Rev. Lett. }}

\def\PRD{{Phys. Rev.  D }}
\def\JAP{{J. Appl. Phys. }}

\def\NPB{{Nucl. Phys.  B }}
\def\NPBP{{Nucl. Phys.  B (Proc. Suppl.) }}

\def\EPJC{{Eur. Phys. J.  C }}
\def\PLB{{Phys. Lett. B }}

\def\PRp{{Phys. Rep. }}
\def\ZPC{{Z. Phys. C }}

\def\JPG{{J. Phys. G }}

\def\APP{{Acta Physica Polon. B }}
\def\AIP{{AIP Conf. Proc. }}

\def\JAA{{J. Astrophys. Astron.}}
\def\noj{{}}

\newcommand{\ee}{\mbox{$e^{+} e^{-}$}\xspace}

\def\Gak{\Gamma\!\left(\!1+\frac{1}{\nu}\right)}
\def\Gbk{\Gamma\!\left(\!1+\frac{2}{\nu}\right)}
\def\Gck{\Gamma\!\left(\!1+\frac{3}{\nu}\right)}
\def\Gdk{\Gamma\!\left(\!1+\frac{4}{\nu}\right)}
\def\Gek{\Gamma\!\left(\!1+\frac{5}{\nu}\right)}

\def\ct{\cite}
\def\beq{\begin{equation}}
\def\eeq{\end{equation}}
\def\bea{\begin{eqnarray}}
\def\eea{\end{eqnarray}}
\def\bear{\begin{array}}
\def\eear{\end{array}}
\def\dsty{\displaystyle}
\def\lbl{\label}
\def\bef{\begin{figure}}
\def\enf{\end{figure}}
\def\capt{\caption}
\def\bec{\begin{center}}
\def\enc{\end{center}}
\def\bib{\bibitem}

\begin{document}

 {\begin{center}
 \Large \bf 
 Describing dynamical fluctuations and genuine correlations 
 by 
 Weibull 
 regularity
 \end{center}
}
\vspace{1cm}

\begin{center}
{\bf 
Ranjit K.~Nayak$^{\rm a,}$\footnote{Email
address: ranjit.nayak@iitb.ac.in},
Sadhana~Dash$^{\rm a,}$\footnote{Email
address: sadhana@phy.iitb.ac.in}, 
Edward K. Sarkisyan-Grinbaum$^{\rm b,c,}$\footnote{Email address:
Edward.Sarkisyan-Grinbaum@cern.ch},
and
Marek Ta\v{s}evsk\'y$^{\rm d,}$\footnote{Email address:
Marek.Tasevsky@cern.ch} 
}
\vspace{0.5cm}


{\it
$^{\rm a}$ 
 Department of Physics, Indian Institute of Technology Bombay,
Mumbai-400076, India\\
$^{\rm b}$ Department of Experimental Physics, CERN, 1211 Geneva 23, 
Switzerland\\
$^{\rm c}$ Department of Physics, The University of Texas at Arlington,
TX
76019, USA\\
$^{\rm d}$ Institute of Physics of the Academy of Sciences of the Czech 
Republic, 18221 Prague, Czech Republic
\\
}

\end{center}

{\bec
\small \bf Abstract
\enc
}
{\small
 The Weibull parametrization of the multiplicity distribution is used to 
describe the multidimensional local fluctuations and genuine multiparticle 
correlations measured by
 OPAL in the large statistics  
 $\ee 
\to Z^0 
\to {
 hadrons}$ sample. 
 The data are found to be well reproduced by the Weibull 
 model 
  up to higher orders. The Weibull predictions are compared to 
 the predictions 
 by 
the two other models, namely by the negative binomial and modified 
negative binomial
   distributions which mostly failed to fit the data.
 The Weibull
 regularity, which is found to 
reproduce
  the multiplicity distributions along with the genuine
  correlations, 
 looks to be the optimal model 
 to describe the multiparticle production process.
}



\section{Introduction}

The 
understanding of multiparticle production in high-energy collisions is 
of 
high ineterest, as, on the one hand, its characteristics are immediate 
observables to provide important information on strong interactions, 
while, on the other hand, this process still faces difficulties in its 
complete understanding by the theory of strong interactions, Quantum 
Chromodynamics (QCD).
 The multiplicity of hadrons produced in high-energy  collisions is one of 
 the 
 most straightforward
measurements and provides a basic information about the dynamics 
of multiparticle production.
 Whereas the full-multiplicity distribution is a global characteristics
 of 
the events and is influenced by the conservation laws, the multiplicities 
averaged over the restricted phase-space regions (bins), or local 
multiplicity fluctuations, are less affected by the global constraints.
 These fluctuations combine contributions of multiparticle correlations 
 which 
 contain crucial information on the details of 
multihadron production mechanisms \ct{book,revp,revi,vhm,multrev}.
 
   The local fluctuations have already attracted high interest and have 
been extensively studied using the method of the
 normalized factorial moments \ct{book,revi}. The moments were proposed to 
extract the 
dynamical, non-Poissonian, fluctuations 
which were
shown to lead to the effect called 
intermittency \ct{intermit}. The intermittency effect shows itself as 
an increase of the 
factorial moments with decrease of the phase-space bin size 
(intermittency scaling).
 The intermittency phenomenon has been searched for in all types of 
collisions -- from leptonic to nuclear ones -- and it was observed to be a 
general feature of the multiparticle production process.
 The characteristics of multihadron production such as (multi)fractality 
(self-similarity) and different phase transition types associated with the 
 intermittency have been widely investigated \ct{book,revp,revi,vhm}. 

 The dynamical fluctuations contain a
 contribution not only of 
 correlations but also from groups of 
 uncorrelated particles. 
 Then, the genuine correlations are mixed up in 
 the fluctuations 
  preventing 
 us 
 to reveal 
the dynamical features of the particle production process.
 To extract the genuine correlations, the method of the 
normalized factorial cumulant 
moments (cumulants) needs to be applied \ct{cums}. 
 The cumulants are constructed from the $q$-particle ($q$th-order) 
cumulant 
correlation functions which vanish whenever one of their arguments becomes 
independent of the others \ct{cummath}, so that they measure the 
{\it genuine} 
correlations. 
 The factorial cumulants remove the influence of the statistical component 
 of the correlations 
   in the same way as the factorial moments do. 
 Employing an advantage of the method to search for {\it many}-particle
correlations, the studies of the high-order ($q\geq 4$) genuine 
correlations have been performed in different types of collisions and 
reveal important features \ct{book,revp,revi}.
 For example, the studies showed that the correlations in nuclear 
collisions are limited by the two-particle correlations while
in {\ee} annihilations 
 the important role of the 
higher-order correlations are 
 obtained,
 the 
 explanation 
 of which 
 was 
 found 
in the many-source dilution 
 effect \ct{dilut}. 
 It was also observed \ct{book} that Monte Carlo models which 
reproduce 
well 
the 
single-particle distributions,  fail to describe the 
measurements of the 
factorial 
moments and cumulants.

 %

 %
 A general belief is that the local fluctuations 
 stem from a generalized random cascade process involved in the particle 
production \ct{bialas} or from branching processes in jet formation 
\ct{ochs}.
 This is why one would expect 
 the intermittency scaling of the moments (and cumulants)  in \ee 
annihilation to be 
 reproduced by the models based on the QCD parton shower with its inherent 
self-similarity. 
 Being 
 a sequential branching process, the parton cascade 
 forms 
 an 
 integral part of the perturbative phase and
   it was also  
 considered  to 
 play a significant 
 role 
 in
 the 
 multiparticle production in \ee interactions.
 However, the QCD-based calculations are found to 
 deviate from the measurements \ct{angul,limpT} indicating a 
 possible 
contribution of 
additional mechanisms 
 to the transition from the 
perturbative partonic phase to the observed 
hadrons. Indeed, the Monte Carlo studies demonstrate \ct{book}  that 
the 
 intermittency
scaling 
manifesting itself 
at the 
level 
of the partonic shower  is much weaker 
than that at the hadronic stage or 
is  smeared out   by the hadronization.

 The branching feature of the intermittency has earlier 
 been 
employed 
\ct{sarkisyan} by applying different parametrizations of the 
multiplicity distributions to the factorial and cumulant 
moments, measured by OPAL
 \ct{opal} in the high-statistics \ee annihilation data sample. 
 The parametrizations which are most popular in multiparticle high-energy 
physics have been used, namely, the negative binomial (NB) distribution, 
its modified (MNB) and generalized forms, the log-normal (LN) 
distribution and the model pure-birth process with immigration (PB).
 Special attention has been devoted to the high-order cumulants ($q$ up to 
5), the only higher-order cumulants measured at LEP.
  It is found that the LN and PB parametrizations fail to describe the 
data, overestimating or underestimating the measurements, respectively. 
 The NB and MNB distributions are found to underestimate the data starting 
from the intermediate-size bins with the MNB spectra found to be closer to 
the measurements.
 The generalized NB distribution is found to show quite a 
complex 
behaviour 
due to different possible 
fitting areas 
of its two parameters, while 
demonstrates the behaviour similar to the NB and LN to which it reduces 
for 
special 
 values
 of the parameters.
 Let us  
 stress 
 that the NB, LN and PB parametrizations are based on one 
parameter which is obtained from the second-order moments. That is, all 
the higher-order moments are based on the two-particle correlations
  and, 
therefore, 
the deviations
 of the predictions 
 indicate the presence and 
importance of high-order correlations.
 This confirms the conclusion made in the measurements \ct{opal} by 
using the 
expansion 
of factorial moments 
by cumulants truncated 
to lower-order terms.
 The MNB has three parameters connected by a condition. In this 
case, the parameters were obtained from the second and third order 
 moments, while 
 the predictions 
 were 
found to underestimate the 
higher-order 
moments 
indicating the inherent high-order correlations in the data.
  
 The parametrizations considered in \ct{sarkisyan} are known to fail to 
 describe the data on the 
 full-phase-space 
 multiplicity distribution  with 
increasing 
energy, and the correlation study 
\ct{sarkisyan} elucidates a reason of this failure.
  For example, the discrepancy between the measured 
high-order correlations and the corresponding 
NB predictions \ct{sarkisyan,opalBEC} points at an inadequacy of 
the single-NB 
to 
describe 
high-multiplicity tails in the multiplicity distribution, so, 
the high-order fluctuations. This would explain a need for 
a combination 
of two or more NB regularities \ct{multiNBD}
or its modification \ct{MNBD} and generalizations \ct{genNBD} to reproduce 
the multiplicity spectra
(see \ct{book,multrev,multrevpp} for reviews).
 
 In this paper, we apply the Weibull parametrization of the multiplicity 
distribution. 
 The Weibull distribution \ct{weib0} is known to be a natural
outcome of complex processes where cascading and sequential branching were 
involved 
 \ct{brown1}.  
 In this sense, one would expect the Weibull distribution to be a natural 
candidate to describe the 
trend of the intermittency scaling. 
 Moreover, in contrast to the parametrizations studied in \ct{sarkisyan}, 
the Weibull 
regularity has been found \ct{hegyi} to 
describe 
well 
the multiplicity 
 distributions 
in different types of collisions and at different energies.
 Recently, it was observed 
 that the Weibull model  
is extremely successful in 
characterizing
 the 
multiplicity distributions 
(in the full phase-space and in its central rapidity intervals) 
up to the top LEP energies in {\ee} annihilations \ct{weib2} 
 and 
up to the 
 TeV LHC energies
 in
 hadronic
 collisions \ct{weib1}. 
  Interestingly, the Weibull distribution is shown \ct{weib2} 
to bridge 
the 
multiplicity distributions in \ee and $p\bar{p}$ interactions pointing 
at  
the multiparticle production universality as the dissipating 
effective-energy approach \ct{disEffE} considers it.  


\section{Weibull distribution, normalized factorial moments and 
cumulants}

 The Weibull distribution is a two-parameter distribution of a random 
variable $n$,

\beq
\nonumber
P(n,\lambda,\nu)=\frac{\nu\, n^{\nu-1}}{\lambda^\nu}
\exp\left[-\left(\frac{n}{\lambda}\right)^\nu\, \right]
\lbl{PWeib}
\eeq
   with the scale parameter $\nu>0$ and the shape parameter $\lambda>0$, 
and 
the mean

\begin{equation}
 \nonumber
 \langle n \rangle = \lambda\, \Gak.
\label{meanW}
\end{equation}
 Here $\Gamma$ denotes the gamma function. The Weibull distribution 
reduces to the exponential distribution for $\nu=1$ and to the  Rayleigh 
distribution for $\nu=2$.   


 For 
 the normalized factorial moments, $F_{q}=\langle 
n(n-1)\cdots (n-q+1)\rangle/{\langle
n \rangle}^q$,
 one finds:
 \bea
 \nonumber
 F_2   = & \!\!
 \dsty  
 \frac{1}{\lambda^2\, {\Gak}^2}& \!\!\! \left\{ 
-\lambda\, \Gak + \lambda^2\, \Gbk \right\}\, ,   \\
 \nonumber
 F_3  = & \!\!   
 \dsty  
 \frac{1}{\lambda^3\, {\Gak}^3}& \!\!\! \left\{ 
 2\, \lambda\, \Gak -3\, \lambda^2\, \Gbk 
 +\lambda^3\, \Gck \right\}\, ,  \\
 \nonumber
 \dsty  
 F_4  = & \!\!    
 \dsty  
 \frac{1}{\lambda^4\,{\Gak}^4}& \!\!\! \left\{
  -6 \lambda\, \Gak + 11 \lambda^2\, \Gbk  
  -6 \lambda^3\, \Gck \right. \\
 \nonumber
 \dsty 
 & & \left. +  \lambda^4\, \Gdk 
 \right\}\, ,  \\
 \nonumber
 F_5  = &\!\!    
 \dsty  
  \frac{1}{\lambda^5\, {\Gak}^5}& \!\!\! \left\{
 24\lambda\, \Gak - 50\lambda^2\, \Gbk 
 +35\lambda^3\, \Gck \right.  \\
 & & \left.
  -  10\lambda^4\, \Gdk 
+\lambda^5\, \Gek  \right\}\, , \;\;\; {\rm etc.} 
 \lbl{FWeib}
 \eea

%
 %
 %

The factorial moments $F_q$ are related to the cumulants $K_q$ through the 
identities \ct{book} (see also \ct{revp,revi,multrev})

\beq
\nonumber
 F_q=\sum_{m=0}^{q-1}
 {{q-1}\choose{m}}
 K_{q-m}\, F_m \, ,
\eeq
 with  
$F_0=F_1=K_1=1$. 
For the first few orders the relations read
 %
\bea
\nonumber
F_2&\!\!\! =&\!\!\! K_2+1\, , \\
\nonumber
F_3&\!\!\! =&\!\!\! K_3+3K_2+1\, ,\\
\nonumber
F_4&\!\!\! =&\!\!\! K_4+4K_3+3K_2^2+6K_2+1\, ,\\
F_5&\!\!\! =& \!\!\! K_5+5K_4+10K_3\,K_2 +10K_3+15K_2^2+10K_2+1\, .
\lbl{FaK}
\eea
 Calculating cumulants $K_q$ one extracts the genuine $q$-particle 
correlations. Note that 
 thanks 
to 
the high statistics of the OPAL data, the 
uniformity of the measured single-particle distributions and the 
corrections 
made 
 in 
\ct{opal}, the translation invariance of the densities 
\ct{cumexp} 
is accounted for in the 
multiplicative terms in the expansions (\ref{FaK}).

\section{Comparison with measurements and discussion}

 \bef[t!]
\includegraphics[scale=0.8]{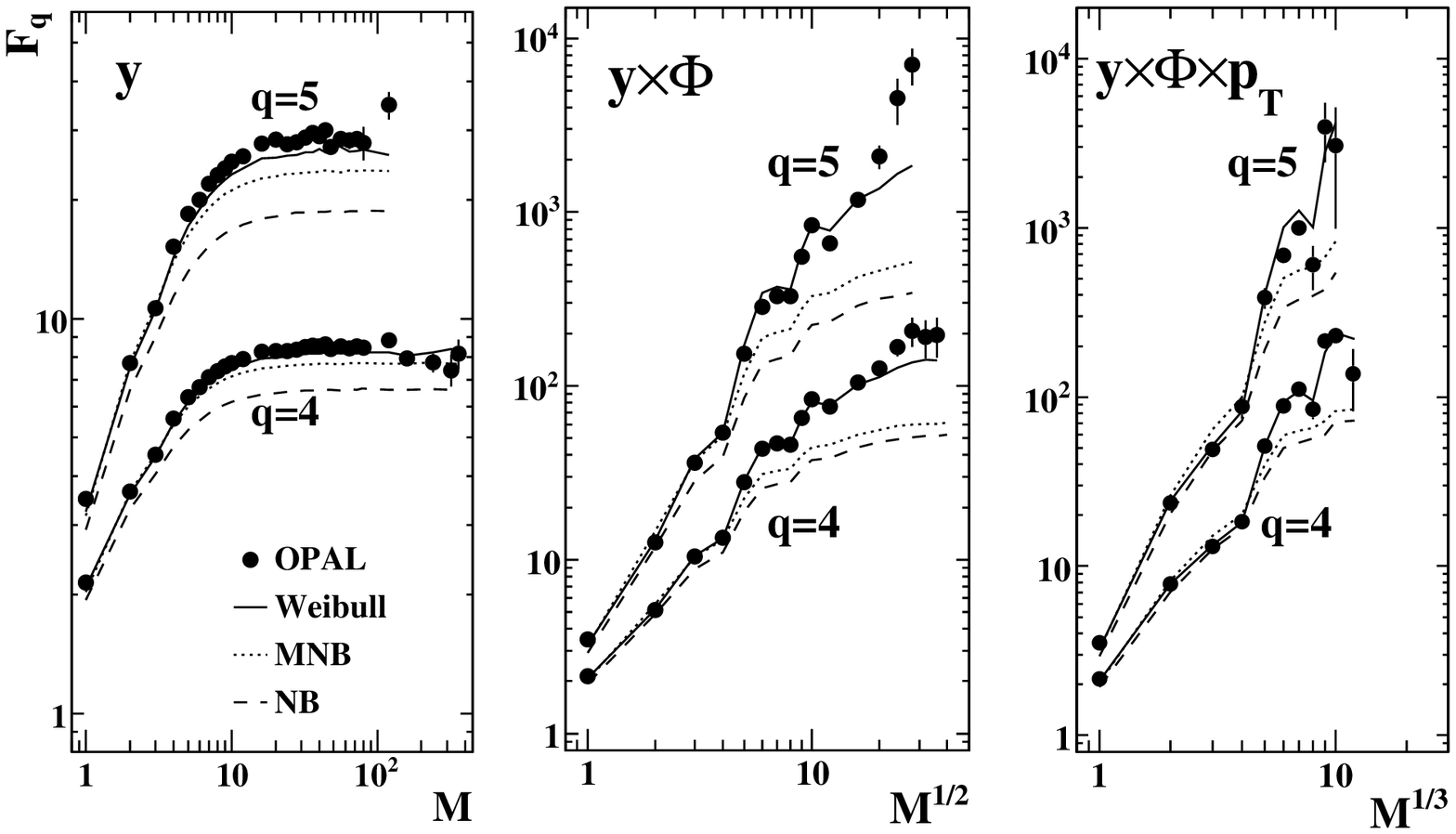} 
\capt{The normalized factorial moments of order $q=4$ and 5 as a 
function of $M^{1/D}$, where $M$ is the number of  bins of the   
 $D$-dimensional subspaces of the  phase space of rapidity, azimuthal  
angle,  and logarithm of transverse momentum, in comparison with the 
predictions of  the Weibull, negative binomial (NB) and modified negative 
binomial (MNB) multiplicity 
parametrizations. The data are taken from \ct{opal}. 
The NB and MNB 
predictions are as in \ct{sarkisyan}.
}
\lbl{fig1}
\enf

\bef[t!]
\includegraphics[scale=0.8]{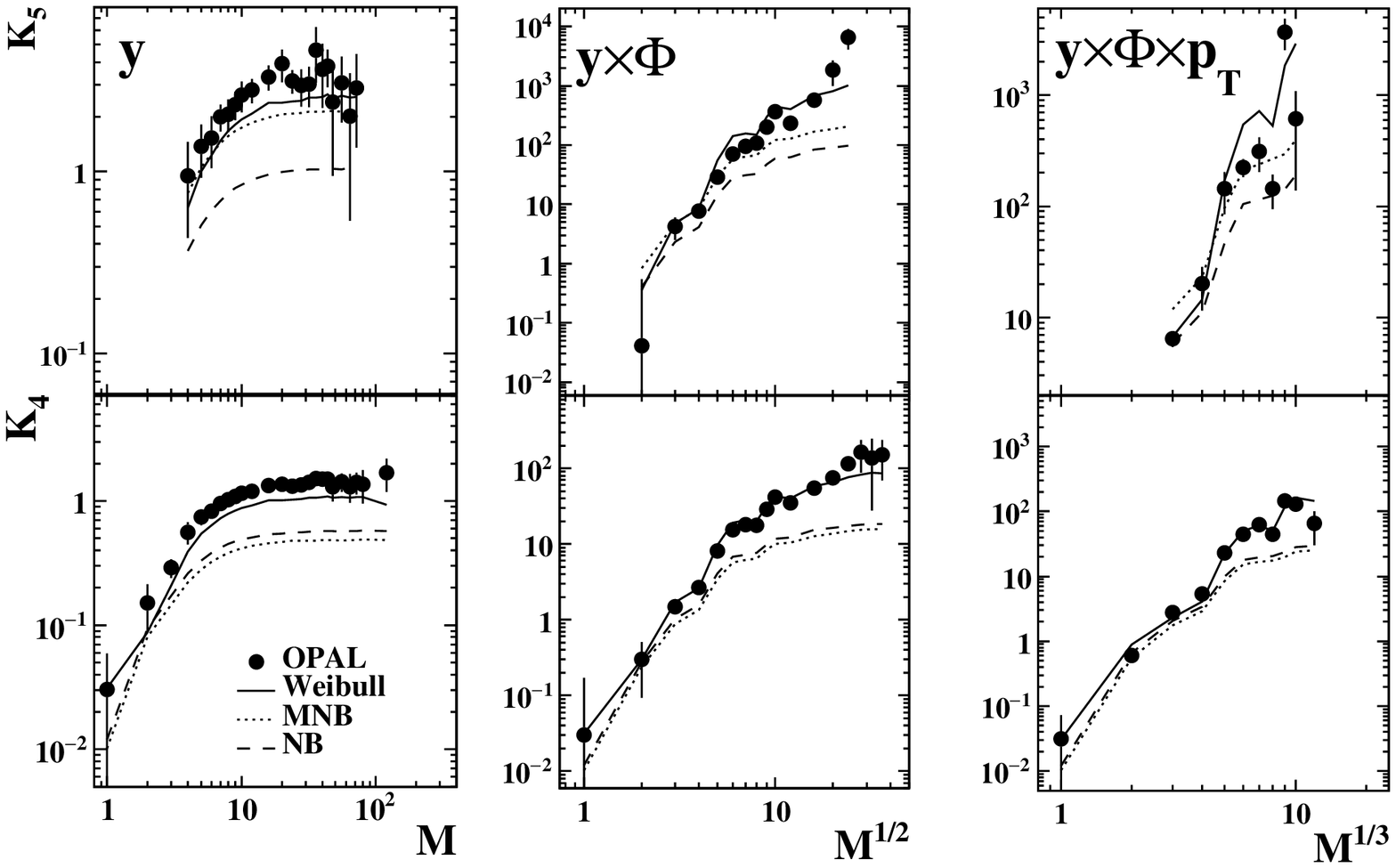} 
\capt{The normalized cumulants of order $q=4$ and 5 as a
  function of $M^{1/D}$, where $M$ is the number of bins of the 
$D$-dimensional subspaces of the  phase space of rapidity, azimuthal  
angle,  
and logarithm of transverse momentum, in comparison with the predictions 
of the  Weibull,
negative binomial (NB) and modified negative binomial (MNB) 
multiplicity parametrizations. The data are taken from \ct{opal}.
The NB and MNB 
predictions are as in \ct{sarkisyan}.
}
\lbl{fig2}
\enf

 Figures 1 and 2 show 
the normalized
factorial moments and the normalized factorial
cumulants, respectively, of the order $q>3$ measured by OPAL \ct{opal} in 
{\ee}
annihilations and compared to a few parametrizations
(lines). 
The
moments are represented in one-, two- and three
dimensions of the phase space of rapidity $y$, logarithm of the 
transverse
momentum $p_T$ and azimuthal angle $\Phi$, calculated with respect
to the sphericity axis.
 In addition to the Weibull predictions,  the NB and 
MNB calculations from 
 \ct{sarkisyan} are also 
given.

The NB predictions \ct{NBD} use the measurements of the normalized 
factorial
moments or cumulants of the second order to compute the NB $k$-parameter, 
$F_2-1=K_2=1/k$,  
and then
the factorial moments and cumulants of order $q\geq 3$ 
are calculated according to the relations 
\beq
\nonumber
F_q=\left(1+\frac{q-1}{k}\right)\, F_{q-1}
\lbl{FNBD}
\eeq
and 
\beq
K_q=(q-1)!\, k^{1-q} \, .
\lbl{KNBD}
\eeq

 From the comparison of 
the NB calculations with the data one can see that 
the 
 calculations 
 underestimate the measurements.
 Moreover, while the calculated $F_q$ moments deviate 
from the data starting from quite large $M$s (see Fig. 1), the cumulants 
clearly show the discrepancy already at small $M$-values (large bin 
sizes), see Fig. 2. The latter confirms 
 that 
 the genuine higher-order 
correlations 
 are 
 present in the data while contributing to the dynamical fluctuations 
along with the lower order correlations, 
 which have been 
extracted by the cumulants in \ct{opal}. 

The MNB 
 regularity leads to 
 the following identities for the cumulants 
\ct{MNBD}:

\beq
\nonumber
K_2=K_2^-+\frac{1}{2\,k\,(r-\Delta)}, \,\,\, 
K_3=K_3^-+\frac{K_2^-}{2\,k\,(r-\Delta)}\, , \,\, {\rm etc.},
\eeq
with
\beq
\nonumber
K_q^-=(q-1)!\, k^{1-q}\, \frac{r^q-\Delta^q}{(r-\Delta)^q} \, ,
\eeq
 where the superscript minus indicates
 cumulants for the negatively charged
particles, and $r=\Delta+\langle  n\rangle /k$.
 The MNB regularity reduces to the NB law if $\Delta=0$, c.f. Eq. 
(\ref{KNBD}). The factorial moments for the MNB regularity are calculated 
using the relations (\ref{FaK}).

The MNB calculations are made based on the third order cumulants in 
the 
sense 
that the parameters $r$ and $\Delta$ have been fixed at the values 
$r=0.91$ 
and $\Delta= - 0.71$, the best values found to describe at least the 
third-order cumulants \ct{sarkisyan}. Then the bin-size dependence is 
regulated by the only 
left parameter $k$, extracted from $K_2$ values. From the results of the 
MNB-based calculations, shown in Figs. 1 and 2, one concludes that the 
 calculations are 
 much 
closer to the data than those made with the NB 
regularity. However, even so, the MNB values underestimate the data for 
large $M$s (small bins), especially in two and three dimensions 
 which 
points to the 
high-order 
 genuine correlations 
 not been large enough 
 in 
 the 
 model.

The values of the normalized factorial moments based on  the Weibull 
distributions, Eqs.(\ref{FWeib}), 
are shown in Fig. 1. The corresponding cumulants, shown in Fig. 2, 
are calculated using the relations (\ref{FaK}). The parameters 
$k$ and $\lambda$ are estimated using the measurements of the second and 
third-order  
factorial moments. 

From Fig. 1 one can see that the Weibull regularity 
describes 
well 
the 
measured factorial moments in one dimension of rapidity and in three 
dimensions, though one might notice discrepancy for the  
largest $M$ for 
$q=5$ in the $y$ subspace. The description is good as well in the 
two-dimensional $y\times \Phi$ subspace except a few extreme $M$ points 
for $q=5$, where the calculations give weaker fluctuations than those 
 observed in the data. 
 One can also see that the Weibull distribution 
gives much better description of the measurements 
compared  to the NB  and MNB distributions.
Similarly, a good description is 
observed for the cumulants, as shown in Fig. 2. From these, one concludes 
that the Weibull regularity well reproduces not only the intermittency 
effect, i.e. the dynamical fluctuations combining genuine correlations of 
different order, but also the genuine high-order correlations.

\section{Summary and conclusions}
 The Weibull distribution  is 
 used 
 to describe the dynamical fluctuations and genuine high-order 
correlations of hadrons in restricted bins of the kinematic phase-space of 
\ee annihilation.
 The study exploits the high-statistics data of multidimensional 
normalized factorial moments and cumulants of charged particles measured 
by OPAL in \ee$\to Z^0$ hadronic decays.
 The Weibull parametrization, which earlier has been shown to well 
 describe the 
 multiplicity distribution in the wide range of the energy, is now found 
to well reproduce the measurements of 
multiplicity fluctuations and correlations 
in restricted bins of the phase space of rapidity, azimuthal angle and 
transverse momentum.
  The Weibull regularity is shown to 
 provide much better
 description of the data compared to 
 that 
 given by other popular regularities such as the negative binomial and  
modified negative binomial distributions which mostly underestimate the 
data.
 This study further establishes that the Weibull distribution, which is 
found to be the first statistical model reproducing simulaneously the 
multiplicity distribution data and the data on the genuine multiparticle 
correlations, looks to be the optimal distribution to describe the 
multiparticle production process.
  The obtained results of the the Weibull calculations to successfully 
describe the crucial characteristics of the multihadron production such as 
genuine correlations, makes it of high interest to be applied to the 
measurements at LHC, where the Weibull model has already been shown to 
reproduce the multiplicity distributions in the full phase-space and in 
the central rapidity intervals.
 This is believed to cast new light on the hadroproduction process and to 
help to elucidate the predictions for the foreseen measurements at LHC, 
FCC and linear colliders.

 %
\bigskip


{\noindent \bf 
Acknowledgements:}
 S.D. thanks the Department of Science and Technology (DST), India, for 
partial support of this work. 
 The work of M.T. is partially supported by the Projects LG15052 and 
LM2015058 of the Ministry of Education of the Czech Republic.

\end{document}